# A Multidimensional Model for Quantifying Tonal Strength: A Continuous Framework of Analyzing Tonal Evolution Beyond Tonal–Atonal Binary Classification

*Yuliang Li*[a,b], *Nan Nan*[a,b,c,*], *Meilian Gu*[d], *and Xiaohong Guan*[a,c,e]
a. Frontier Institute of Science and Technology, Xi'an Jiaotong University, Xi'an, China;
b. Interdisciplinary Research Center of Frontier Science and Technology, Xi'an Jiaotong University, Xi'an, China;
c. MOE KLINNS Lab, Faculty of Electronic and Information Engineering, Xi'an Jiaotong University, Xi'an, China;
d. Department of Music Artificial Intelligence and Music Information Technology, Central Conservatory of Music, China;
e. Center for Intelligent and Networked Systems, Tsinghua University, Beijing, China.
* Corresponding author: Nan Nan. Email: nan.nan@xjtu.edu.cn

## Abstract

Tonality is a fundamental organizing principle of Western music. However, the gradual transition from common-practice tonality to atonality around the turn of the twentieth century cannot be captured by binary labels. To characterize intermediate tonal states and trace this evolution, we propose a multidimensional framework for quantifying tonal strength, grounded in tonal theory and mechanisms underlying tonal dissolution. This framework operationalizes the global statistical regularity, temporal coherence, and local perceptual alignment of tonal organization through three complementary features: pitch-class distribution uniformity, temporal stability of tonal centers, and local tonal clarity, respectively. To validate the model, we introduce two new datasets: a binary dataset comprising 196 tonal and atonal pieces, and a historical MIDI corpus of 1,561 compositions by 21 composers spanning the Baroque era to the early twentieth century. The proposed model achieves high discriminative performance on the binary classification task ($F_1 = 0.985$), and quantitatively reveals a continuous historical trajectory from established tonality through progressive weakening to eventual dissolution. Computational analysis of representative composers further confirms significant declines in tonal strength throughout the careers of late-Romantic and modernist figures. Additionally, analysis of EMOPIA dataset reveals a weak but statistically significant positive correlation between higher tonal strength values and positive emotional valence. This work establishes a unified quantitative model for analyzing tonal evolution and provides a continuous measure for future behavioral and neurophysiological investigations of the relationships among tonal organization, musical emotion, and cognition.

## 1. INTRODUCTION

Tonality is one of the most fundamental organizing principles in Western music. Since the emergence of the common-practice tradition, hierarchical relationships among pitch classes, chords, and keys have provided a framework through which listeners perceive musical structure, stability, and expectation. In music theory, tonality is commonly understood as the orientation of musical events toward a referential tonal center, typically represented by the tonic pitch class and its associated harmonic functions (Hyer, 2002). This tonal organization underlies much of the Western classical repertoire and remains a central concept in music analysis, music cognition, and music information retrieval (MIR). Empirical studies in music cognition have further demonstrated that listeners internalize stable tonal hierarchies through exposure, as reflected in psychologically derived tonal templates. For instance, Krumhansl & Kessler (1982) quantified the perceived stability of pitch classes in Western major and minor tonalities.

Since tonality plays a crucial role in shaping musical structure and emotion, it has long attracted attention in computational musicology and MIR research. Many computational approaches use tonality-related features for tasks such as key estimation, harmonic analysis, music similarity measurement, style characterization, and corpus exploration. Among the representations used to derive such features, pitch-class distributions (PCDs) are particularly important because they summarize how pitch classes are distributed within a musical work and are closely related to perceived key and harmonic organization (Lieck et al., 2020). The use of PCDs is closely connected to early cognitive models of tonal perception, which showed that aggregated pitch-class frequencies strongly correlate with perceived tonal centricity (Krumhansl, 1990). PCDs can therefore serve as a common representation from which complementary tonal features are derived at both global and local levels. Such features provide valuable information for studying long-term stylistic developments in Western music history and for comparing musical repertoires across historical periods (Harasim et al., 2021; Weiß et al., 2019).

Beyond these structural applications, recent work in music cognition suggests that tonal organization is also related to affective perception. Empirical studies have reported associations between tonal hierarchy and perceived emotional valence, with more structurally stable tonal configurations tending to elicit more positive affective judgments (Maimon et al., 2022). This relationship motivates examining whether a continuous computational measure of tonal strength can capture systematic variation in emotional valence.

Traditional computational approaches to tonality have primarily focused on key identification. The influential Krumhansl–Schmuckler algorithm estimates the most likely key by comparing observed pitch-class distributions with experimentally derived tonal templates (Krumhansl & Kessler, 1982). Subsequent work extended this paradigm through probabilistic models (Temperley, 2007). More recent research has examined local key estimation across recordings and annotators (Weiß et al., 2020), developed deeper convolutional architectures and broader data-augmentation strategies for global key estimation (Baumann, 2021), and introduced musically informed convolutional–recurrent models for more robust local key estimation (Ding & Weiß, 2024). Although these methods have improved the identification and temporal tracking of tonal centers, their primary outputs remain global or local key labels. They therefore do not directly provide a continuous measure of the overall strength of tonal organization.

This limitation becomes particularly evident when analyzing music from the late nineteenth and early twentieth centuries. During this period, Western art music underwent a gradual transformation from common-practice tonality toward increasingly chromatic and eventually atonal idioms. The transition was not abrupt but instead involved numerous intermediate compositional languages, including extended tonality, modal writing, impressionism, and free atonality. Consequently, many works cannot be adequately described using a binary distinction between tonal and atonal music. Rather, they occupy different positions along a continuum of tonal strength. Similar concerns have motivated recent corpus-based studies that investigate style evolution through continuous musical features instead of discrete historical categories (Henzel et al., 2025; Moss et al., 2024). Theoretical accounts of tonal prolongation and hierarchical pitch-space structure suggest that tonal organization can be understood in terms of graded structural relationships rather than solely discrete key categories (Lerdahl, 2001).

Several researchers have attempted to move beyond binary classification. For example, Honingh & Bod (2010) proposed using pitch-class-set categories to estimate degrees of tonality and demonstrated that certain set classes occur substantially more often in tonal than in atonal repertoires. While such approaches provide valuable insights, they typically rely on a single statistical perspective and therefore capture only one aspect of tonal organization. Tonality, however, is a multidimensional phenomenon involving pitch-class distribution uniformity, temporal stability of tonal centers, and local tonal clarity.

To address this gap, we propose a multidimensional model for quantifying tonal strength in symbolic music based on tonal theory. Throughout this paper, tonal strength refers to the underlying continuous property of tonal organization, whereas the tonal-degree value ($T$) denotes the numerical estimate of this property produced by the proposed model. The model operationalizes tonal strength through three complementary features: (1) pitch-class distribution uniformity, which measures how closely the composition-level pitch-class distribution approximates a uniform reference distribution; (2) temporal stability of tonal centers, which captures the temporal continuity of local tonal organization; and (3) local tonal clarity, which measures the degree of correspondence between local pitch distributions and tonal templates. These features are integrated into a unified framework to estimate the tonal-degree value ($T$).

To evaluate the proposed approach, we constructed two datasets: a binary classification corpus of 196 works comprising tonal piano works and atonal works by the Second Viennese School composers, and a symbolic corpus of 1,561 works by 21 composers spanning the Baroque period to the early twentieth century. Using the former, we validate the efficacy of the proposed model in distinguishing tonal from atonal music. Using the latter, we employ the proposed model to characterize the continuous historical gradient of diminishing tonality from the common-practice period to modernism, and further demonstrate that the resulting tonal-degree values can effectively capture stylistic evolution trajectories across individual composers' careers.

The contributions of this paper are fourfold. First, we introduce a theoretically interpretable multidimensional framework for modeling tonal strength. Second, we construct and curate two complementary symbolic-music datasets that support both tonal–atonal benchmark evaluation and large-scale analysis of historical tonal evolution. Third, we demonstrate that the proposed model provides highly accurate discrimination between tonal and atonal repertoires. Finally, we demonstrate the broader analytical utility of the resulting tonal-degree values through two complementary applications: tracing long-term and intra-composer tonal evolution and providing initial corpus-level evidence of a weak but statistically reliable association between tonal strength and emotional valence in EMOPIA.

The remainder of this paper is organized as follows. Section 2 reviews related work on key estimation, tonal–atonal classification, tonal strength modeling, and emotional valence. Section 3 describes the datasets and corpus design, and Section 4 presents the proposed multidimensional model of tonal strength. Section 5 reports tonal–atonal classification results, historical and intra-composer analyses, and the relationship between tonal-degree values and emotional valence. Sections 6 and 7 discuss the findings and conclude the paper, respectively.

## 2. RELATED WORK

### 2.1 Key Estimation

Tonal recognition refers to the task of identifying the most likely key of a musical excerpt, typically formulated as a classification problem over major and minor keys (e.g., C major, A minor). One of the most influential cognitive foundations for this task is the work of Krumhansl & Kessler (1982), who derived experimentally obtained tonal templates through probe-tone experiments, resulting in 24 key-specific pitch-class stability distributions. These profiles form the basis of the Krumhansl–Schmuckler (K–S) algorithm, which estimates tonal center by correlating observed pitch-class distributions with reference profiles. Subsequent improvements include probabilistic formulations and time-adaptive extensions (Temperley, 2007), which improve robustness in modulation contexts.

Beyond cognitive-inspired models, modern computational approaches have combined feature engineering with data-driven learning. Pitch-class-based representations, such as chroma features and the Harmonic Pitch Class Profile (HPCP), remain widely used in audio-based key estimation because of their relative robustness to timbral variation (Gómez, 2006). More recent learning-based systems derive task-specific harmonic representations from audio-based time–frequency representations. Building on earlier end-to-end CNN approaches (Korzeniowski & Widmer, 2017), a deeper Inception-based architecture combined with broad data augmentation has improved global key-estimation performance (Baumann, 2021), while a musically informed convolutional–recurrent network has shown stronger generalization in local key estimation (Ding & Weiß, 2024). Vision-transformer architectures have also been evaluated for musical key estimation; however, a recent comparison found that DenseNet achieved higher overall accuracy than the evaluated transformer models, although the transformer variants performed better on some temporal metrics (Garg et al., 2023).

Despite their success, these methods primarily aim to predict a discrete key label, and their internal representations are optimized for classification accuracy rather than explicit modeling of tonal strength or structural stability.

### 2.2 Tonal–Atonal Classification and Tonal Strength Quantification

A second line of research focuses on distinguishing tonal and atonal music, as well as quantifying tonal strength. Early work by Honingh & Bod (2010) introduced pitch-class set (PCS) categories (IC1–IC6) as analytical tools for tonal organization, showing that tonal music contains significantly higher proportions of diatonic set classes (e.g., IC5 corresponding to diatonic collections) compared to atonal music. This framework demonstrated that tonal structure can be partially captured through statistical distributions of pitch-class sets.

Another related approach is interval-based clustering and classification, which uses interval distributions to separate tonal and atonal repertoires (Honingh & Bod, 2011), achieving strong discrimination performance between tonal and atonal corpora. However, these methods remain fundamentally categorical in nature, focusing on classification rather than continuous representation.

More recent corpus-based studies have traced stylistic evolution using time-varying quantitative tonal descriptors rather than relying solely on discrete period labels. Based on 2,000 audio recordings spanning more than 300 years, Weiß et al. (2019) constructed evolution curves for chord-transition patterns, interval usage, and tonal complexity, showing that gradual feature changes and data-driven groupings broadly corresponded to conventional historical periodization. Using a large historical corpus, Moss et al. (2024) inferred piece-level distributions over six directed interval relations in tonal space—ascending and descending perfect fifths, major thirds, and minor thirds—and identified increasing tonal-interval complexity, the long-term predominance of fifth relations, and a marked rise in third relations during the nineteenth century associated with extended tonality. These studies demonstrate the value of continuous corpus analysis for tracing historical stylistic change; however, they were designed primarily to characterize specific diachronic trends rather than to provide a general, interpretable framework for representing tonal strength continuously and comparably across individual works, composers, and historical periods.

Overall, existing key-estimation methods focus on identifying discrete key labels, classification approaches distinguish tonal from atonal music, and corpus-based studies trace selected tonal features over historical time. However, these approaches do not provide a unified, multidimensional, and interpretable representation of tonal strength as a continuous variable.

### 2.3 Tonality and Emotional Valence

Recent research in music cognition has provided evidence for a systematic association between tonal structure and perceived emotional valence. In particular, hierarchical relationships among scale degrees have been shown to correlate with affective judgments, such that more structurally stable or "closural" tonal elements tend to be associated with more positive emotional valence. Importantly, this association has been observed in both musicians and non-musicians, suggesting that tonal–affective mappings may arise from general perceptual mechanisms rather than musical expertise alone (Maimon et al., 2022).

However, this relationship is not fully reducible to traditional measures of tonal goodness-of-fit or harmonic stability. Empirical comparisons between tonal hierarchy judgments and affective association tasks suggest that tonal valence may reflect partially distinct cognitive processes, including both explicit, knowledge-driven evaluation and implicit, automatic associations. Related work further supports the view that tonal closure and key-related structure can influence perceived emotional content across development and listening contexts (Suberry et al., 2020).

Together, these findings suggest that tonal organization may carry affective implications beyond its structural function, motivating computational approaches that quantify tonal strength as a continuous variable when studying music perception and emotion.

## 3. DATASET AND CORPUS DESIGN

To evaluate the proposed model for quantifying tonal strength, we construct two complementary datasets designed to capture both categorical discrimination and continuous stylistic variation in tonal organization. This dual design reflects the central objective of this study: modeling tonality as a continuous structural property rather than a discrete label.

### 3.1 Binary Tonal–Atonal Evaluation Dataset

The first dataset consists of 196 symbolic MIDI files, including 100 tonal piano excerpts selected from the EMOPIA (Hung et al., 2021) dataset and 96 symbolic representations drawn from the atonal and post-tonal repertoire of composers of the Second Viennese School.

The EMOPIA dataset consists of contemporary piano performances written within the tonal tradition and therefore serves as the tonal subset in our evaluation. The dataset provides a clear contrast between music exhibiting stable tonal organization and music characterized by highly chromatic or post-tonal harmonic language.

The purpose of this dataset is not to perform broad stylistic classification, but to evaluate whether the proposed model can reliably distinguish between two fundamentally different forms of tonal organization.

### 3.2 Historical Corpus for Continuous Tonal Analysis

The second dataset is a large-scale historical corpus consisting of 1,561 symbolic compositions by 21 Western composers, spanning from the Baroque period to the early twentieth century. Table 1 summarizes the historical corpus used for evolutionary analysis. The data were collected from GitHub, MAESTRO dataset (Hawthorne et al., 2019), kunstderfuge (https://kunstderfuge.com) and MuseScore (https://musescore.com).

Unlike the binary dataset, this corpus is not intended for classification, but for continuous analysis of tonal change. It enables the investigation of whether tonal strength exhibits systematic temporal trends both across historical periods and within individual composers' creative trajectories.

### 3.3 Design Rationale

The combination of the two datasets reflects two complementary evaluation perspectives. The binary dataset provides a controlled setting for validating the discriminative capacity of the proposed model, while the historical corpus supports analysis of tonal organization as a continuous variable over time.

This design aligns with the central hypothesis of this work: tonal structure should be modeled not only as a categorical distinction between tonal and atonal music, but also as a continuous and evolving property across musical history. Together, these datasets provide the empirical foundation for both validation and longitudinal analysis of tonal strength.

## 4. METHOD

In this section, we propose a multidimensional framework for quantifying tonal strength of symbolic Western music. Rather than treating tonality as a binary category (tonal vs. atonal), we model it as a continuous structural property emerging from multiple interacting aspects of pitch organization.

| Historical period | Representative composer | Data source | Number of pieces |
|---|---|---|---|
| Baroque | J.S. Bach (1685–1750) | GitHub | 135 |
| | D. Scarlatti (1685–1757) | MAESTRO | 29 |
| Classical | J. Haydn (1732–1809) | MAESTRO | 29 |
| | W.A. Mozart (1756–1791) | MAESTRO | 76 |
| | L.v. Beethoven (1770–1827) | MAESTRO | 95 |
| Early and mid-Romantic | F. Schubert (1797–1828) | MAESTRO | 96 |
| | F. Mendelssohn (1809–1847) | kunstderfuge | 58 |
| | F. Chopin (1810–1849) | MAESTRO | 107 |
| | R. Schumann (1810–1856) | MAESTRO | 31 |
| | F. Liszt (1811–1886) | MAESTRO | 91 |
| | J. Brahms (1833–1897) | MAESTRO | 23 |
| Late Romantic and post-Romantic | P. Tchaikovsky (1840–1893) | kunstderfuge | 274 |
| | S. Rachmaninoff (1873–1943) | MAESTRO | 50 |
| | R. Wagner (1813–1883) | MuseScore | 52 |
| | G. Mahler (1860–1911) | MuseScore | 118 |
| | R. Strauss (1864–1949) | MuseScore | 78 |
| Impressionism and modernist transition | C. Debussy (1862–1918) | MAESTRO | 41 |
| | A. Scriabin (1872–1915) | MAESTRO | 26 |
| Second Viennese School | A. Schoenberg (1874–1951) | MuseScore | 68 |
| | A. Webern (1883–1945) | MuseScore | 65 |
| | A. Berg (1885–1935) | MuseScore | 19 |

**Table 1.** Summary of the historical composer corpus for tonal-evolution analysis

Grounded in music-theoretical and cognitive accounts of tonality, the proposed method conceptualizes tonal structure as a hierarchical system centered on a tonic and internalized through perceptual exposure (Krumhansl, 1990; Krumhansl & Kessler, 1982; Lerdahl & Jackendoff, 1983).

We quantify tonality through three complementary dimensions: pitch-class distribution uniformity ($U$), temporal stability of tonal centers ($S$), and local tonal clarity ($C$). These features jointly capture multiple aspects of tonal organization and are integrated into a single tonal-degree value.

### 4.1 Pitch-Class Distribution Uniformity ($U$)

Tonal music exhibits structured and non-uniform pitch-class distributions that reflect hierarchical functional organization, whereas atonal music tends toward more uniform distributions (Schoenberg, 1975).

Let each composition be represented by a normalized pitch-class duration vector:

$$\mathbf{p} = (p_1, p_2, \dots, p_{12}), \sum_{i=1}^{12} p_i = 1, \tag{1}$$

where $p_i$ denotes the proportion of the composition's total accumulated note duration assigned to pitch class $i$, with $i \in \{1, \dots, 12\}$ indexing the 12 chromatic pitch classes. Thus, $\mathbf{p} \in \mathbb{R}^{12}$, $p_i \geq 0$, and $\sum_{i=1}^{12} p_i = 1$. Then we define a uniform reference distribution:

$$\mathbf{u} = \left(\frac{1}{12}, \frac{1}{12}, \dots, \frac{1}{12}\right), \tag{2}$$

where $\mathbf{u} \in \mathbb{R}^{12}$ denotes the uniform reference distribution and $u_i = 1/12$ is the reference probability assigned to each pitch class.

Pitch-class distribution uniformity ($U$) is defined as the cosine similarity between the pitch-class vector and the uniform reference:

$$U = \frac{\mathbf{p} \cdot \mathbf{u}}{|\mathbf{p}||\mathbf{u}|}, \tag{3}$$

where $\mathbf{p} \cdot \mathbf{u}$ denotes the dot product of the two vectors, $|\mathbf{p}|$ and $|\mathbf{u}|$ denote the Euclidean norm, and $U$ is the pitch-class distribution uniformity value. Higher values of $U$ indicate a pitch-class distribution closer to uniformity, which corresponds to weaker tonal strength. Conversely, lower values indicate stronger deviation from uniformity and thus stronger tonal structure. For consistency with the other two features (where higher values indicate stronger tonality), we apply a direction alignment in Section 4.5.

### 4.2 Temporal Stability of Tonal Centers ($S$)

Tonal perception involves the continuous inference of tonal centers over time, yielding a coherent hierarchical interpretation of musical structure (Lerdahl & Jackendoff, 1983). To quantify this temporal dimension, each composition is segmented into overlapping analysis windows $w_t$, each comprising four notes, with a step size of two notes (50% overlap). The four-note window length was selected with reference to Krumhansl (1990), who showed that the K-S algorithm correctly identified the keys of 44 of Bach's 48 preludes using only their first four notes; the robustness of this window length is further evaluated in Section 5.1.

For each window $w_t$, a local tonal center $k_t$ is estimated by computing the correlation between the window's pitch-class distribution and 24 standard tonal templates (Krumhansl & Kessler, 1982; Temperley, 1999):

$$k_t = \arg\max_{k \in \mathcal{K}} \operatorname{corr}(\boldsymbol{w}_t, \boldsymbol{t}_k), \tag{4}$$

where $t \in \{1, \dots, N\}$ indexes the analysis windows, $N$ is the total number of windows in the composition, $\boldsymbol{w}_t \in \mathbb{R}^{12}$ is the normalized pitch-class duration vector of window $t$, $\mathcal{K}$ is the set of 24 major and minor keys, and $\boldsymbol{t}_k \in \mathbb{R}^{12}$ is the tonal-template profile associated with key $k$. The function $\operatorname{corr}(\cdot,\cdot)$ denotes Pearson's correlation coefficient, and $k_t \in \mathcal{K}$ is the key whose template has the highest correlation with window $t$.

We define the temporal stability as the mean similarity between successive windows:

$$S = \frac{1}{N-1} \sum_{t=1}^{N-1} \operatorname{corr}(\boldsymbol{t}_{k_t}, \boldsymbol{t}_{k_{t+1}}), \tag{5}$$

where $S$ is the temporal-stability value, $N \geq 2$ is the total number of analysis windows, and $\boldsymbol{t}_{k_t}$ and $\boldsymbol{t}_{k_{t+1}}$ are the tonal templates corresponding to the estimated tonal centers of two successive windows. Thus, $S$ is the mean Pearson correlation between the tonal-center templates of all adjacent window pairs.

This feature measures the temporal stability of the tonal center. As shown in Figure 1, stronger tonal organization is reflected by a relatively stable high-value region in the heat map, with only small changes across adjacent windows. The visualization follows the representation proposed by Toiviainen & Krumhansl (2003). Accordingly, higher values of ($S$) indicate that the inferred tonal center remains relatively stable over time, whereas lower values indicate more frequent or larger shifts in the inferred tonal center, reflecting weaker temporal coherence of tonal organization. As illustrated in Figure 1, the tonal excerpt follows the sequence {C major, C major, C major, G minor}, with successive tonal-template correlations of 1.000, 1.000, and 0.243, yielding ($S = 0.748$). By contrast, the atonal excerpt follows the sequence {G minor, B minor, E minor, C major}, with successive correlations of −0.005, 0.340, and 0.536, yielding ($S = 0.290$). This quantitative contrast demonstrates that ($S$) captures differences in the temporal persistence of inferred tonal centers between the two examples.

### 4.3 Local Tonal Clarity ($C$)

Tonal cognition research has demonstrated that perceived tonal strength depends on the alignment between local pitch content and internalized tonal schemas (Bigand et al., 2014). For each analysis window $w_t$, we compute the maximum correlation between the window's pitch-class distribution and the set of 24 tonal templates:

$$C_t = \max_{k \in \mathcal{K}} \operatorname{corr}(\boldsymbol{w}_t, \boldsymbol{t}_k), \tag{6}$$

where $C_t$ is the local tonal-clarity value of window $t$. All other symbols are defined as in Eq. (4). A larger $C_t$ indicates a stronger correspondence between the pitch-class distribution of window $t$ and its best-matching tonal template. The composition-level local tonal clarity $C$ is then defined as the mean maximum correlation across all windows:

$$C = \frac{1}{N} \sum_{t=1}^{N} C_t, \tag{7}$$

where $C$ is the composition-level local tonal-clarity value, $C_t$ is the local tonal clarity of window $t$, and $N$ is the total number of analysis windows. Thus, $C$ represents the mean maximum template correlation across all windows in the composition. This feature measures how strongly local segments conform to canonical tonal structures. Higher values indicate that the composition consistently aligns with well-defined tonal centers at the local level. As shown in Figure 1, local tonal clarity is reflected in the activation pattern within each individual window: a concentrated region of high activation suggests a clearly defined local tonal center, whereas a more diffuse activation pattern indicates greater tonal ambiguity. Quantitatively, the maximum correlations with the best-fitting tonal templates across the four windows are 0.883, 0.866, 0.689, and 0.655 for the tonal excerpt, yielding ($C = 0.773$). The corresponding correlations for the atonal excerpt are 0.423, 0.370, 0.503, and 0.478, yielding ($C = 0.444$). The higher ($C$) value of the tonal excerpt therefore confirms its clearer correspondence with established tonal templates.

### 4.4 Rationale for Multidimensional Integration

Although the three features $U$, $S$, and $C$ capture distinct aspects of tonal strength, they are theoretically interrelated. Music theory describes tonality as a unified hierarchical system rather than a set of independent properties (Lerdahl, 2001; Schenker, 1935). From a cognitive perspective, perceptual studies have shown strong correlations between pitch distribution, tonal stability, and perceived tonal strength, suggesting a shared underlying tonal schema.

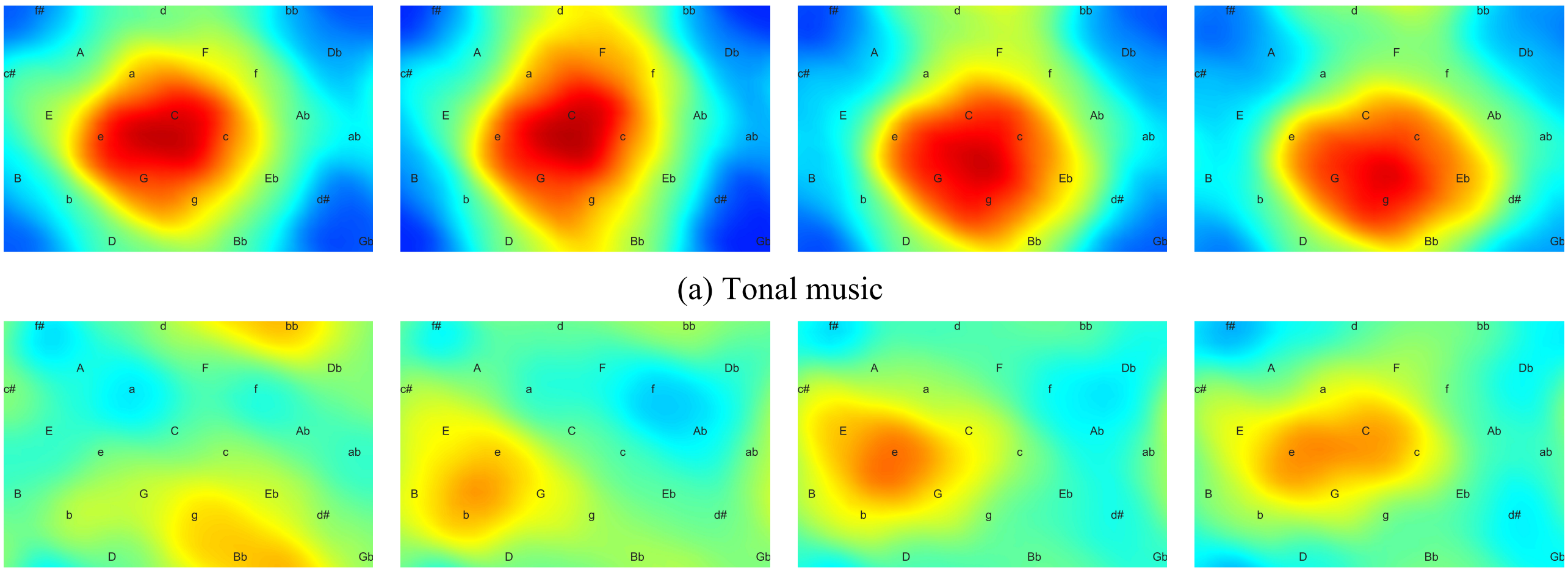

(a) Tonal music

(b) Atonal music

**Figure 1.** Temporal evolution of local tonal-template correlation heatmaps in representative tonal and atonal excerpts. Rows show (a) tonal music and (b) atonal music; within each row, the four heatmaps correspond sequentially to Windows 1–4. Each heatmap shows the correlations between the local pitch-class distribution and the 24 major and minor tonal templates. Red indicates higher correlation and a more clearly defined local tonal center, whereas blue indicates lower correlation and greater tonal ambiguity. All panels use the same color scale.

We therefore treat the three observed features as complementary indicators of tonal strength and integrate them into a continuous tonal-degree value. Let the observed feature vector be:

$$\mathbf{x} = [U', C, S]^T, \tag{8}$$

where $U' = 1 - U$ is the direction-aligned version of the pitch-class distribution uniformity measure $U$, such that higher values of all three features correspond to stronger tonality.

### 4.5 Multidimensional Feature Integration

We integrate the direction-aligned features $U'$, $C$ and $S$ into a single tonal-degree value $T$. The feature vector is standardized as $\mathbf{z} = (\mathbf{x} - \boldsymbol{\mu})/\boldsymbol{\sigma}$, where $\boldsymbol{\mu}$ and $\boldsymbol{\sigma}$ are the corresponding sample means and standard deviations. The integration is implemented using factor-analytic estimation, which provides a transparent linear mapping from the observed features to the final model output:

$$\mathbf{z} = \boldsymbol{\lambda} T + \boldsymbol{\epsilon}, \tag{9}$$

where $T$ is the tonal-degree value, $\boldsymbol{\lambda} = [\lambda_1, \lambda_2, \lambda_3]^T$ is the vector of factor loadings, and $\boldsymbol{\epsilon} = [\epsilon_1, \epsilon_2, \epsilon_3]^T$ is the vector of unique variances (noise terms). We assume:

$$T \sim \mathcal{N}(0,1), \epsilon_i \sim \mathcal{N}(0, \psi_i), \tag{10}$$

where $T$ and $\boldsymbol{\epsilon}$ are independent. This leads to the following covariance structure for the observed features:

$$\boldsymbol{\Sigma} = \boldsymbol{\lambda}\boldsymbol{\lambda}^T + \boldsymbol{\Psi}, \tag{11}$$

where $\boldsymbol{\Psi} = \mathrm{diag}(\psi_1, \psi_2, \psi_3)$ is the diagonal covariance matrix of the unique variances.

**Interpretation.** The factor loading $\lambda_i$ measures the contribution of the latent tonality $T$ to the observed feature $x_i$. A higher loading indicates that the feature is more strongly determined by the underlying tonal construct. The unique variance $\psi_i$ captures variance specific to each feature that is not shared with the latent dimension.

**Model estimation.** The parameters $\boldsymbol{\lambda}$ and $\boldsymbol{\Psi}$ are estimated from the sample covariance matrix $\mathbf{S}$ using maximum likelihood estimation (Jöreskog, 1967). For identification, we set the variance of $T$ to 1 and fix the sign of one loading (we use $\lambda_1 > 0$) to resolve rotational ambiguity.

### 4.6 Tonal-Degree Value Computation

After model estimation, the tonal-degree value for each composition is obtained as the estimated latent factor score from the fitted maximum-likelihood factor analysis model. The estimated latent factor score is mathematically equivalent to the classical regression factor score (Thomson, 1951), which can be expressed as:

$$\hat{T} = \mathbf{w}^T \mathbf{z}, \mathbf{w} = \hat{\boldsymbol{\Sigma}}^{-1} \hat{\boldsymbol{\lambda}}, \tag{12}$$

where $\hat{\boldsymbol{\Sigma}} = \hat{\boldsymbol{\lambda}}\hat{\boldsymbol{\lambda}}^T + \hat{\boldsymbol{\Psi}}$ is the model-implied covariance matrix, and $\hat{\boldsymbol{\lambda}}, \hat{\boldsymbol{\Psi}}$ are the maximum likelihood estimates. Substituting Eq. (12), this becomes:

$$\hat{T} = \hat{\boldsymbol{\lambda}}^T (\hat{\boldsymbol{\lambda}}\hat{\boldsymbol{\lambda}}^T + \hat{\boldsymbol{\Psi}})^{-1} \mathbf{z}, \tag{13}$$

where all symbols are defined as in Eqs. (9)-(12). The resulting tonal-degree value $\hat{T}$ is a continuous measure centered at zero in the estimation sample, where higher values correspond to stronger tonal strength. For consistency of value magnitude and theorical tradition, we verify the sign alignment such that the correlation between $\hat{T}$ and the direction-aligned features is positive.

**Algorithm 1** Computation of the tonal-degree value

**Input:** Symbolic music file ($F$); analysis window size ($W$); hop size ($H$); scaler parameters (($\boldsymbol{\mu}, \boldsymbol{\sigma}$)); estimated factor-model parameters (($\boldsymbol{\lambda}, \boldsymbol{\Psi}$)).
**Output:** Tonal-degree value ($T$).

```
Parse (F) and extract note events.
Compute the global pitch-class duration distribution (p).
Compute the pitch-class distribution uniformity (U).
Apply direction alignment: (U' ← 1 − U).
for each analysis window (w_t) do
    Compute the local pitch-class distribution.
    Estimate the local tonal center (k_t).
    Compute the local tonal clarity (C_t).
end for
Compute temporal stability of tonal centers (S).
Compute the local tonal clarity (C).
Form the feature vector (x = [U', C, S]^T).
Standardize the feature vector: (z = (x − μ)/σ).
Estimate the tonal-degree value (T) using Eq. (13).
```

### 4.7 Summary

The proposed model provides a theoretically grounded, multidimensional approach to quantifying tonal strength in symbolic music. It combines three complementary features: pitch-class distribution uniformity ($U$), temporal stability of tonal centers ($S$), and local tonal clarity ($C$). After direction alignment, ($U'$), ($S$), and ($C$) are integrated into a single tonal-degree value ($T$), where higher values indicate stronger tonal strength and lower values indicate weaker tonal strength.

By combining pitch-class distribution uniformity, temporal stability of tonal centers, and local tonal clarity, the proposed model captures multiple aspects of tonal strength within one continuous measure. This makes the model suitable not only for tonal–atonal classification, but also for large-scale corpus analysis of tonal change across composers and historical periods.

## 5. EXPERIMENTS AND RESULTS

The binary classification experiment uses the 196-piece tonal–atonal evaluation dataset, whereas the historical and intra-composer analyses use the 1,561-piece historical corpus; the emotional-valence analysis uses EMOPIA.

### 5.1 Experimental Setup

Short-term tonal perception operates on the scale of a few notes or about one second (Krumhansl, 1990). We therefore set the analysis window to $W = 4$ notes, which in our corpus corresponds to approximately one beat in common meter, matching the timescale of local tonal impression formation (Huron & Parncutt, 1993). To evaluate the robustness of the proposed method with respect to analysis-window size, we varied $W$ from 4 to 20 notes. As shown in Figure 2, the local tonal clarity feature consistently achieved AUC values above 0.99 across all tested window sizes, while the temporal stability of tonal centers feature maintained AUC values between 0.947 and 0.965, indicating only minor performance variation. These results demonstrate that the proposed features are robust to reasonable variations in analysis-window size and do not require extensive parameter tuning. Accordingly, all subsequent experiments use $W = 4$ with a hop size of $H = 2$ (50% overlap).

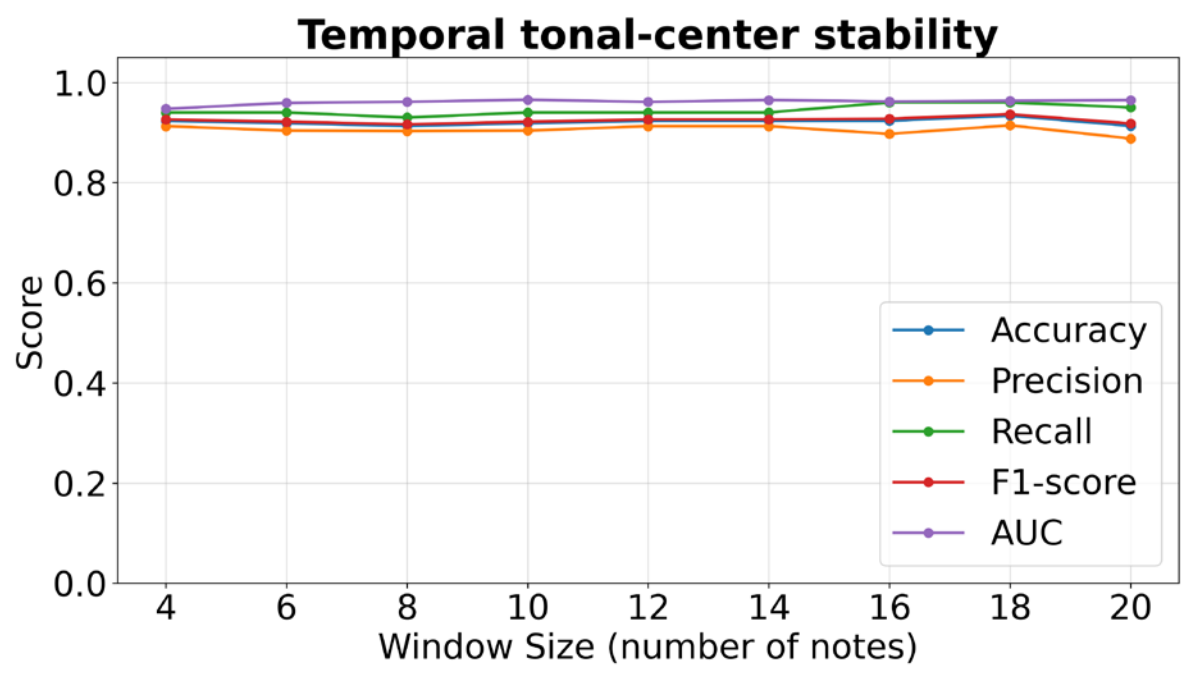

(a) Temporal tonal-center stability

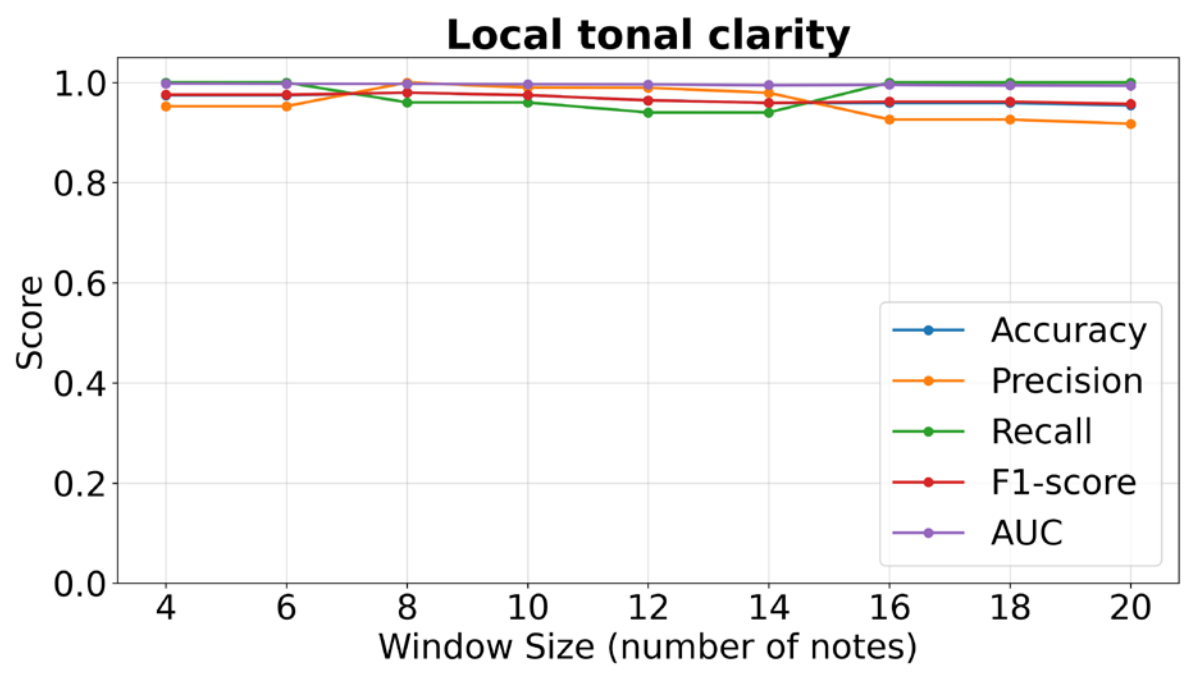

(b) Local tonal clarity

**Figure 2.** Classification performance of the proposed tonal features under different analysis-window sizes.

Before constructing the integrated tonal-degree value $T$, we evaluated whether the three tonal features shared sufficient common structure for one-dimensional integration. As shown in Table 2, the KMO value was 0.756, and Bartlett's test of sphericity was significant, indicating sufficient shared structure among the features. The first principal component had an eigenvalue of 2.644 and explained 87.67% of the total variance, motivating the subsequent single-factor model.

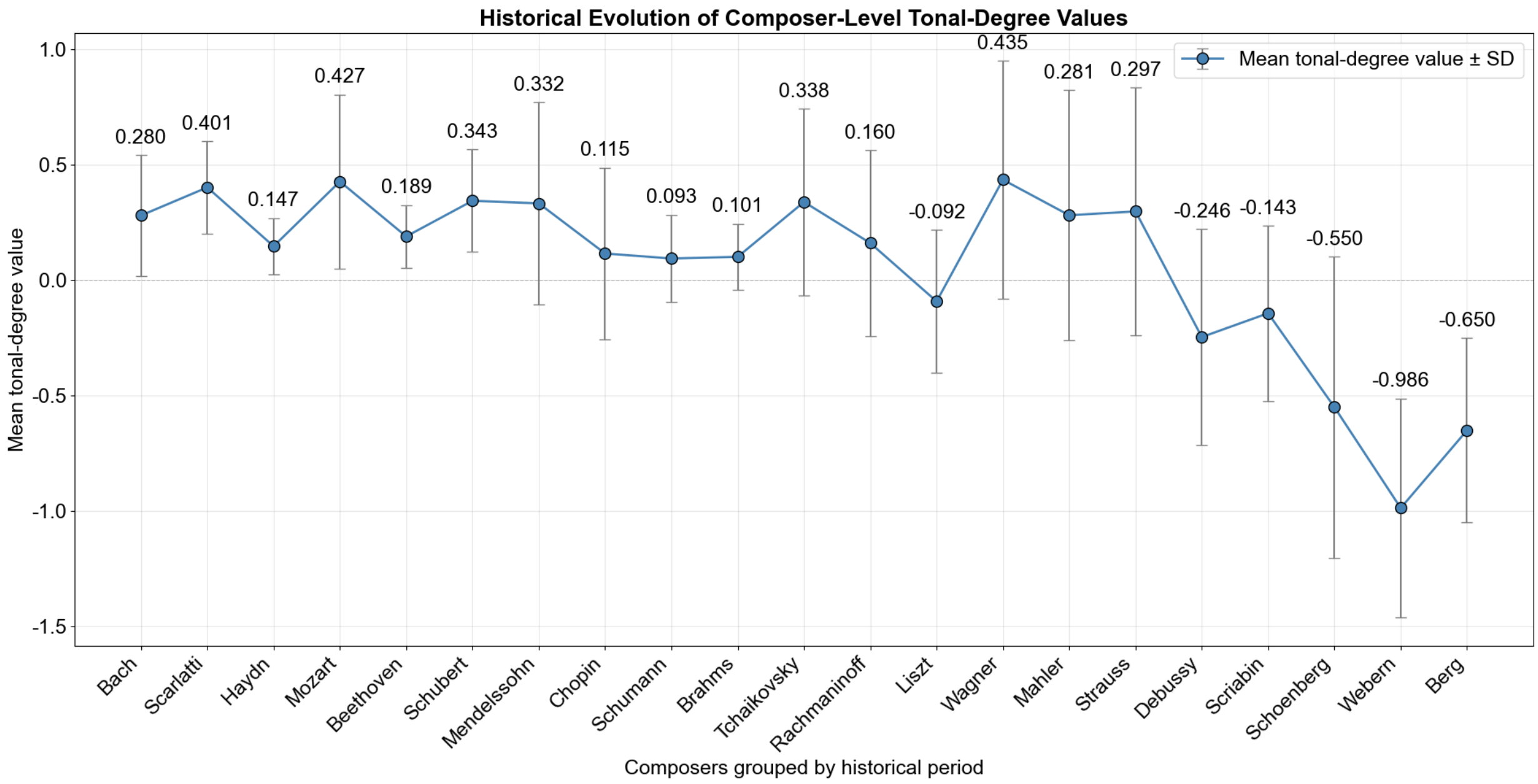

**Figure 3.** Historical evolution of composer-level tonal-degree values.

| Diagnostic measure | Result |
|---|---|
| KMO measure of sampling adequacy | 0.756 |
| Bartlett's test of sphericity | $\chi^2 = 475.77$, $p < 0.001$ |
| First eigenvalue | 2.644 |
| Variance explained by the first principal component | 87.67% |
| Number of retained factors | 1 |

**Table 2.** Diagnostics for one-dimensional feature integration.

As an external baseline, we implement a windowed adaptation of the alternative classification algorithm proposed by Honingh & Bod (2011). Each composition is segmented into 24-note windows with a hop size of 12 notes. Within each window, the algorithm identifies the diatonic scale containing the largest number of notes and counts the notes lying outside that scale.

The mean number of out-of-scale notes across all analysis windows is then used as the classification score, with pieces exceeding a decision threshold classified as atonal and the remainder as tonal.

We also investigated the interval-category-based method presented in the same study. However, despite repeated implementation attempts, we were unable to reproduce the reported performance from the information provided in the publication. Therefore, to ensure a fair and reproducible comparison, we report results only for the alternative algorithm.

For each evaluated feature or method, we selected a decision threshold that maximized the $F_1$ score on the 196-piece benchmark. For the integrated tonal-degree value, the sample-optimal cutoff was 0.06; scores at or above this cutoff were classified as tonal, and lower scores as atonal.

### 5.2 Binary Classification Results

On the 196-piece benchmark, the proposed model achieves near-perfect classification performance at the sample-optimal cutoff of 0.06 ($F_1 = 0.985,\ AUC = 0.999$), outperforming all individual features and the alternative algorithm baseline (Table 3).

### 5.3 Historical Evolution of Tonality

Figure 3 plots the mean tonal-degree value for each composer, with composers arranged according to historical-period and stylistic groupings; error bars represent ±1 standard deviation. This visualization illustrates both the overall historical trajectory of tonal strength and the distribution of tonal-degree values across the sampled works of each composer. Here, zero denotes the center of the estimation sample and serves as a reference for descriptive comparison.

Baroque and Classical composers all have positive means, and their lower bounds (mean $-\ \sigma$) remain above zero, indicating that the bulk of their works are firmly within the tonal region, consistent with the stable tonal syntax of the common-practice period (Rosen, 1997).

| Method | Accuracy | Precision | Recall | $F_1$ | AUC |
|---|---|---|---|---|---|
| Pitch-class distribution uniformity | 0.9541 | 0.9417 | 0.9700 | 0.9557 | 0.9841 |
| Local tonal clarity | 0.9745 | 0.9524 | **1.0000** | 0.9756 | 0.9977 |
| Temporal stability of tonal centers | 0.9235 | 0.9126 | 0.9400 | 0.9261 | 0.9472 |
| Alternative classification algorithm (Honingh & Bod, 2011) | 0.9745 | 0.9798 | 0.9700 | 0.9749 | 0.9966 |
| Proposed multidimensional model | **0.9847** | **0.9802** | 0.9900 | **0.9851** | **0.9985** |

**Table 3.** Tonal–atonal classification performance.

In the Romantic period, several composers (Chopin, Schumann, Brahms) still have positive means, but their lower bounds dip below zero, reflecting that individual works within their output already show weakened tonal strength, reflecting the increasing use of chromaticism, distant modulations, and delayed resolutions (Plantinga, 1984). Liszt is the first Romantic to show a negative mean (−0.09), indicating that some of his works—especially those from his late period—depart markedly from traditional tonality.

Debussy and Scriabin have negative means, but their upper bounds ($\text{mean} + \sigma$) remain positive. This reveals that even though their overall output is weakly tonal, some works still retain traces of tonal center—consistent with music-historical descriptions of impressionist and mystic harmony (Simms, 1986).

The Second Viennese School shows the strongest departure. Schoenberg's mean is negative (−0.55) and his upper bound is slightly above zero, reflecting the wide stylistic range from his early tonal works to his twelve-tone period. Berg and Webern, however, have both negative means and negative upper bounds, indicating that their entire compositional output falls predominantly in the atonal region.

Webern's mean (−0.99) is the lowest among the sampled composers, indicating the most pronounced weakening of tonal strength in the corpus (Simms, 1986).

Collectively, these findings reveal a continuous gradient: from uniformly positive values with lower bounds > 0 (Baroque/Classical), through positive means with negative lower bounds (Romantic), to negative means with positive upper bounds (Debussy, Scriabin, Schoenberg), and finally to fully negative intervals (Berg, Webern). This quantitative spectrum aligns closely with music-historical narratives of tonal evolution.

### 5.4 Intra-Composer Analysis

To examine whether the model output varies systematically within individual composers, we tested the association between composition year and tonal-degree values for Wagner, Mahler, Strauss, and Schoenberg. Figure 4 visualizes these intra-composer temporal patterns, with each point representing one sampled work and the dashed line indicating the fitted linear trend. As shown in Table 4, all four composers show negative associations in the sampled corpus, and Mahler and Schoenberg reach statistical significance. These results are treated as corpus-based statistical patterns of the model output, rather than as definitive claims about each composer's stylistic development.

| Composer | slope | $p$ value | Significance |
|---|---|---|---|
| Mahler | −0.0225 | 0.001** | ** |
| Schoenberg | −0.0232 | 0.012* | * |
| Strauss | −0.0052 | 0.070 | n.s. |
| Wagner | −0.0074 | 0.373 | n.s. |

**Table 4.** Statistical association between composition year and tonal-degree values.

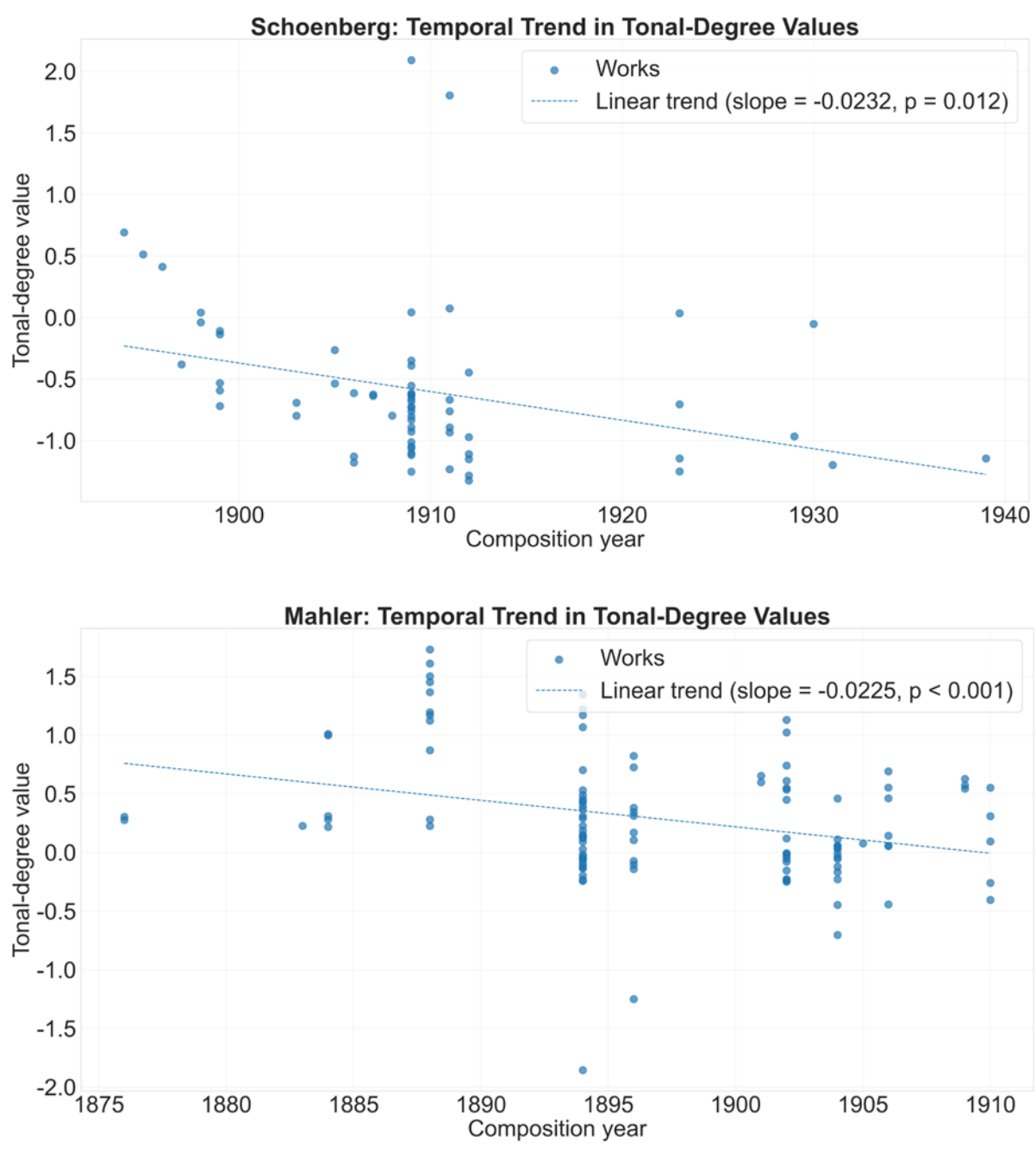

**Figure 4.** Intra-composer temporal trends in tonal-degree values.

Schoenberg's steep decline traces his evolution from late-Romantic tonality (Verklärte Nacht, 1899, score near zero) through free atonality (Op. 11, 1908, ca. −1.01) to twelve-tone technique (e.g., Op. 33a, around −0.7). Musicologists (Haimo, 1990) describe this as a staged break with tonality; our model quantifies it as a continuous gradient, with the large standard deviation (0.652) reflecting the stylistic span of his career.

Mahler shows a significant decline, consistent with observations that his later symphonies progressively destabilize tonality (Hefling, 1997). Strauss and Wagner show no significant trend despite large standard deviations, aligning with the view that both composers expanded tonality without dismantling it.

### 5.5 Tonal-Degree Values and Emotional Valence in EMOPIA

To evaluate the relationship between tonal structure and emotional valence, we analyzed tonal-degree values in the EMOPIA dataset. In this analysis, Q1 and Q4 were grouped as positive-valence music, while Q2 and Q3 were grouped as negative-valence music.

Positive-valence excerpts show higher tonal-degree values ($M = 1.033, SD = 0.319, N = 555$) compared to negative-valence excerpts ($M = 0.965, SD = 0.359, N = 516$).

A Mann–Whitney $U$ test reveals a statistically significant difference between the two groups ($U = 155450, p = 0.00768$, one-sided), indicating that tonal-degree values are higher in positive-valence music than in negative-valence music.

The effect size is small ($r = 0.074$), suggesting a weak but statistically reliable association between tonal structure and emotional valence.

## 6. DISCUSSION

### 6.1 Interpretation of the Multidimensional Model

The effectiveness of the individual features suggests that tonal organization is reflected in multiple, partially overlapping indicators. The purpose of multidimensional integration is not to make tonal–atonal discrimination possible, but to combine these indicators so that the model does not depend on any single aspect of tonal organization. This is particularly relevant to transitional or ambiguous works, in which different tonal properties may not change in parallel. A composition may preserve clear local tonal references while frequently shifting tonal centers, or retain a structured global pitch distribution without consistent local alignment with tonal templates. Considering these patterns jointly provides a more balanced representation of such mixed tonal states.

The shared structure among the features supports their integration along a common dimension, but does not imply that they are interchangeable. In this framework, the tonal-degree value represents a continuous composite estimate based on converging tonal evidence, rather than a categorical key label or a replacement for the individual features. The integrated value provides a common scale for comparing graded differences across works and repertoires, while the component features remain useful for interpreting the particular dimensions underlying those differences.

### 6.2 Music-Historical Implications

The historical patterns identified by the model broadly accord with musicological accounts of stable common-practice tonality, nineteenth-century chromatic expansion, and the early twentieth-century dissolution of traditional tonal organization (Rosen, 1997; Plantinga, 1984; Simms, 1986). More importantly, the overlap and dispersion of tonal-degree values across periods and composers suggest that this transition is better understood as a heterogeneous continuum rather than as a sequence of discrete stylistic boundaries.

The continuous historical pattern does not replace conventional period labels, which remain useful for contextualizing musical style. Instead, the tonal-degree value reveals variation that these labels may obscure, including differences among works from the same period and changes within an individual composer's output. The observed chronological gradient should therefore be interpreted as a corpus-level tendency rather than as a deterministic developmental sequence shared by every composer or composition.

### 6.3 Implications for Music-Emotion Research

Tonal strength is one of several musical factors related to emotional valence. Emotional interpretation is also influenced by mode, tempo, register, rhythm, dynamics, and expressive performance. These factors may interact with tonal strength. In this study, the tonal-degree value provides a continuous estimate of tonal strength. It should not be treated as a standalone predictor of emotion. Instead, it can be included as a continuous covariate in multivariate analyses or used to select experimental stimuli with controlled levels of tonal strength. This measure could also support future behavioral and neurophysiological studies examining how variations in tonal strength relate to listeners' emotional judgments under controlled experimental conditions.

### 6.4 Limitations

The current study is limited to symbolic MIDI representations of Western art music from the Baroque to the early twentieth century. Extending the model to audio recordings and to other musical traditions (e.g., jazz, popular music, non-Western tonal systems) would test its generalizability. Additionally, the current integration procedure assumes a linear mapping from the observed features to the resulting tonal-degree value; future work could explore nonlinear alternatives.

## 7. CONCLUSION

This study proposed a multidimensional framework for quantifying the tonal strength of Western music, operationalized through three complementary and music-theoretically grounded features: pitch-class distribution uniformity, temporal stability of tonal centers, and local tonal clarity. Through a principled linear factorization approach, these features are integrated into a continuous and interpretable tonal-degree value that captures gradations of tonal organization across historical repertoires.

To evaluate the framework, we constructed two complementary symbolic-music datasets: a 196-piece benchmark for tonal–atonal discrimination and a historical corpus of 1,561 works by 21 composers spanning the Baroque period to the early twentieth century. Together, these datasets support evaluation of both categorical discrimination and continuous tonal variation across composers and historical periods.

Across four complementary analyses, the model achieved high tonal–atonal classification performance, captured both historical and intra-composer variation in tonal strength, and provided preliminary evidence of an association with emotional valence in EMOPIA. Grounded in tonal theory and empirical evaluation, the framework translates qualitative descriptions of tonality into a reproducible quantitative measure, bridging traditional music theory and computational musicology. It also provides a quantitative basis for music cognition research and for the parametric design of experimental stimuli in future behavioral and neurophysiological studies of tonal perception.

## 8. DATA ACCESSIBILITY

The complete implementation of the proposed model is publicly available at:

https://anonymous.4open.science/r/Tonality-Degree-Model-GitHub

The repository includes all feature-extraction procedures, the factor-analytic integration framework, the MIDI corpora, processed tonal features, and the analysis scripts used for binary classification, longitudinal corpus analysis, and emotion-valence analysis, thereby facilitating reproducibility of the reported results.

## 9. COMPETING INTERESTS

The authors have no competing interests to declare.